\documentclass[journal]{IEEETran}
\usepackage{amsmath,graphicx,cite,epsfig}
\usepackage{multirow}
\usepackage{booktabs}
\usepackage{float}
\usepackage{subfigure}
\usepackage{caption}
\usepackage{threeparttable}
\usepackage{diagbox}
\usepackage{url}
\usepackage{bm}
\usepackage{stfloats}
\usepackage{flushend}

\ifCLASSINFOpdf

\else

\fi

\begin{document}
%
\title{Reduced-Reference Quality Assessment of Point Clouds via Content-Oriented Saliency Projection}
%
%
%

\author{Wei Zhou, Guanghui Yue, Ruizeng Zhang, Yipeng Qin, and Hantao Liu
\thanks{This work was supported in part by NSFC under Grant 62001302 and Guangdong Basic and Applied Basic Research Foundation 2019A1515111205.}
\thanks{W. Zhou is with the Department of Electrical and Computer Engineering, University of Waterloo, Waterloo, ON N2L 3G1, Canada (e-mail: wei.zhou@uwaterloo.ca).}
\thanks{G. Yue is with the School of Biomedical Engineering, Shenzhen University, Shenzhen 518060, China (e-mail: yueguanghui@szu.edu.cn).}
\thanks{R. Zhang is with Beijing Institute of Technology, Beijing 100081, China (e-mail: hireason@163.com).}
\thanks{Y. Qin and H. Liu are with the School of Computer Science and Informatics, Cardiff University, Cardiff, CF24 4AX, United Kingdom (e-mail: qiny16@cardiff.ac.uk; liuh35@cardiff.ac.uk).}
\thanks{}}

\maketitle

\begin{abstract}
Many dense 3D point clouds have been exploited to represent visual objects instead of traditional images or videos. To evaluate the perceptual quality of various point clouds, in this letter, we propose a novel and efficient Reduced-Reference quality metric for point clouds, which is based on Content-oriented sAliency Projection (RR-CAP). Specifically, we make the first attempt to simplify reference and distorted point clouds into projected saliency maps with a downsampling operation. Through this process, we tackle the issue of transmitting large-volume original point clouds to user-ends for quality assessment. Then, motivated by the characteristics of the human visual system (HVS), the objective quality scores of distorted point clouds are produced by combining content-oriented similarity and statistical correlation measurements. Finally, extensive experiments are conducted on SJTU-PCQA and WPC databases. The experimental results demonstrate that our proposed algorithm outperforms existing reduced-reference and no-reference quality metrics, and significantly reduces the performance gap between state-of-the-art full-reference quality assessment methods. In addition, we show the performance variation of each proposed technical component by ablation tests.
\end{abstract}


%
\IEEEpeerreviewmaketitle

\section{Introduction}
%
%
%
%
\IEEEPARstart{M}{assive} visual data are emerging in our daily lives. Especially in recent years, due to the rapid development of 3D capture and rendering technologies, point cloud has become one of the most popular immersive media formats. It leads to lots of real-world applications, such as automatic diving, mixed reality, remote sensing, and so on \cite{guo2020deep,schwarz2018emerging}. Usually, a dense point cloud is a 3D model and has a set of scattered points in space, which is employed to represent an object. Each point owns geometry coordinates (i.e., $x$, $y$, $z$) and photometric attributes (e.g., RGB color information). Similar to conventional images/videos, point clouds have also undergone a variety of artifacts during the multimedia signal processing chain, from acquisition, compression, to transmission, reconstruction, and display \cite{sun2022graphiqa,chikkerur2011objective,zhou2022blind}. In other words, these procedures will inevitably cause quality degradation at user-ends. Therefore, the quality evaluation of point clouds is crucial to monitor and guarantee satisfactory user quality-of-experience.

Since humans are the ultimate receivers of point clouds, subjective quality assessment is the most reliable method. By carrying out relevant subjective tests, several point cloud quality databases have been established \cite{yang2020predicting,liu2022perceptual,alexiou2019comprehensive,javaheri2020point}. However, such subjective tests are often time-consuming, expensive, and labor-intensive. Thus, how to design effectively objective quality assessment methods based on the human visual system (HVS) is a challenging yet promising research direction.

Generally, for the objective quality metrics of point clouds, there exist three categories, including full-reference (FR), no-reference (NR), and reduced-reference (RR). When pristine information is entirely accessible, FR methods are proposed by comparing the difference or similarity between the distorted and the corresponding original reference point clouds. The earliest FR metrics are used for the standardization body of MPEG point cloud compression, which consist of point-to-point \cite{mekuria2016evaluation}, point-to-plane \cite{tian2017geometric}, point-to-mesh \cite{cignoni1998metro}, and plane-to-plane \cite{alexiou2018point}, etc. These methods compute the distance deviation between reference and distorted point clouds. Afterward, Yang et al. \cite{yang2020inferring} proposed the graph similarity index (GraphSIM) method based on graph signal processing, and also extended it to a multiscale variant \cite{zhang2021ms}. Inspired by the structural similarity index (SSIM) \cite{wang2004image}, Alexiou et al. \cite{alexiou2020towards} exploited geometry, normal vectors, curvature values, and color information to form PointSSIM. Moreover, in \cite{meynet2019pc}, Meynet et al. used local curvature statistics to construct PC-MSDM and further proposed the point cloud quality metric (PCQM) \cite{meynet2020pcqm} according to the optimally weighted linear combination of curvature and color features. Lu et al. \cite{lu2022point} compared the 3D edge similarity to quantify point cloud quality. Except for direct comparisons on 3D point cloud models, another kind of FR methods is to project point clouds into multiple 2D images from different views. Then, the mainstream 2D quality evaluation algorithms can be utilized, such as PSNR and SSIM.

In real scenarios, the full information of original reference point clouds may not be available. Several NR methods are developed to predict the visual quality from distorted point clouds, both from hand-crafted and learning-based aspects. For example, Zhang et al. \cite{zhang2022no} proposed NR-3DQA for evaluating point cloud quality on the basis of 3D natural scene statistics (NSS). In \cite{liu2021pqa}, a point cloud quality assessment network (PQA-Net) was designed by multi-task learning. In addition, by view projection, 2D quality metrics in NR manner can be applied, e.g., the representative NSS methods called blind/referenceless image spatial quality evaluator (BRISQUE) \cite{mittal2012no} and natural image quality evaluator (NIQE) \cite{mittal2012making}. Suppose when part of reference information can be obtained, RR metrics provide an intermediate solution. Nevertheless, very few RR quality assessment methods for point clouds have been proposed in the literature. Liu et al. \cite{liu2021reduced} exploited quantization parameters to estimate
the perceptual quality of V-PCC compressed point clouds. To the best of our knowledge, there is only one general-purpose RR metric named PCMRR \cite{viola2020reduced} which relies on a small set of statistical features from the original point clouds regarding geometry, normal vector, and color. The characteristics of the HVS are lacking in the design philosophy of the PCMRR method.

To fill the blank of the above-mentioned problem, we propose the RR metric based on Content-oriented sAliency Projection (RR-CAP). The main contributions of this letter are summarized as follows:
\begin{enumerate}
    \item Motivated by the HVS properties, we propose the first image-based RR point cloud quality assessment method via saliency projection.
    \item The content-oriented similarity and statistical correlation measurements are developed in the proposed framework.
    \item Our proposed quality metric can relieve the transmission burden for the large amount of point data, while still show competitive performance when tested on subject-rated databases.
\end{enumerate}

The rest of this letter is organized as follows. Section II first introduces the proposed RR-CAP method in detail. In Section III, we then show the experimental results and analyses on publicly available point cloud quality databases. Finally, the conclusion is given in Section IV.


\section{Proposed RR-CAP Metric}
\begin{figure*}[t]
	\centerline{\includegraphics[width=17.5cm]{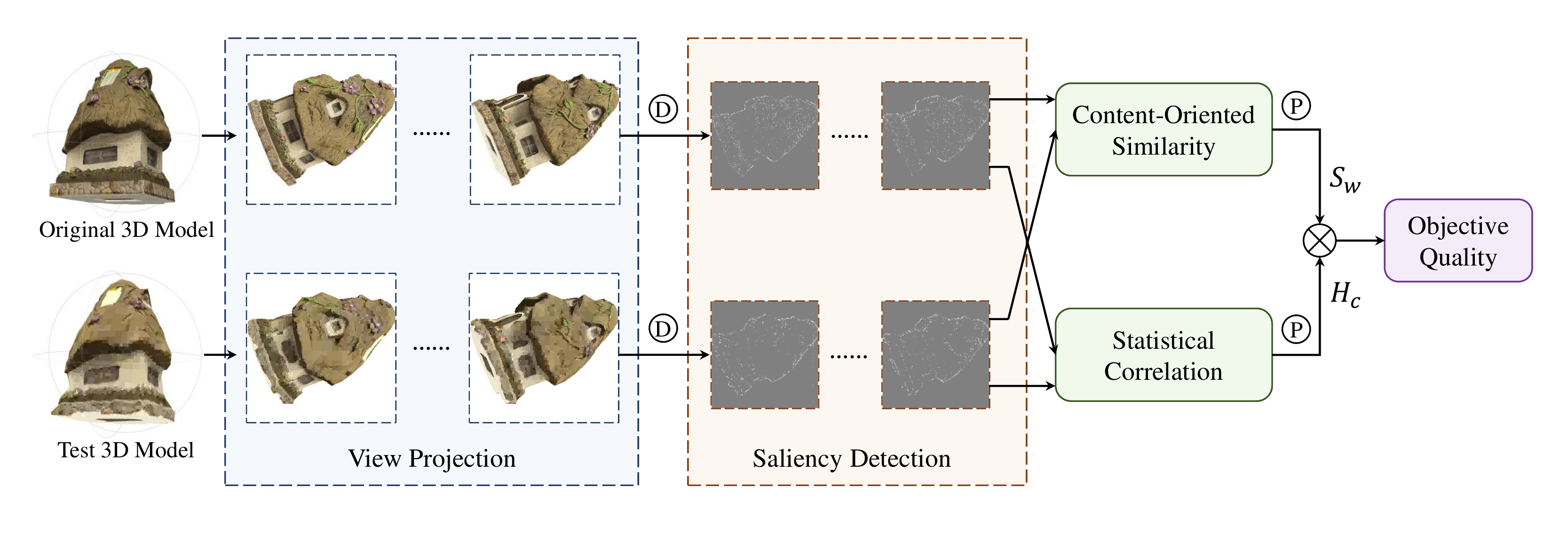}}
	\caption{Overall framework of the proposed RR-CAP, where $\normalsize{\textcircled{\scriptsize{D}}}\normalsize$ and $\normalsize{\textcircled{\scriptsize{P}}}\normalsize$ are the downsampling and pooling operations, respectively.}
	\centering
	\label{fig1}
\end{figure*}

The overall framework of our proposed RR-CAP is shown in Fig. \ref{fig1}. To reduce the large-scale point cloud data from reference side, we first extract downsampled saliency maps after view projection. The content-oriented similarity and statistical correlation are then combined to estimate the perceptual quality of test point cloud.

\subsection{Saliency Extraction in Projected Planes}
When subjects browse the targeted point clouds, they would synthesize the quality sensation from all views, resulting in the final experience. Therefore, we project both original and test 3D models into multiple 2D image planes. Here, we adopt the six perpendicular projections \cite{yang2020predicting} which can uniformly cover most of the viewed content.

Because introducing artifacts may change the saliency behaviors and the HVS is more sensitive to distortions in salient areas, visual saliency plays a vital role in quality evaluation \cite{lin2022visual,min2018saliency}. After view projection, we extract saliency maps by image signature \cite{hou2011image}. Specifically, for each projected image $I$, we first downsample it to a coarser counterpart by:

\begin{equation}
\tilde{I}(i, j)=A * I(s i, s j),
\end{equation}
where $s$ represents the downsampling scale. $i=1,2, \ldots, \frac{I}{s}$ and $j=1,2, \ldots, \frac{J}{s}$ are indexes for the row and column of downsampled image, respectively. $A$ is a low-passing filter and $*$ denotes the convolution operation.

Then, the image signature is defined as the sign function of discrete cosine transform (DCT) coefficients \cite{saad2012blind} for downsampled projection, which can be computed as follows:

\begin{equation}
\hat{I}=\operatorname{sign}(\operatorname{D C T}(\tilde{I})).
\end{equation}
With the image signature in the transformed format, we can convert it back to the spatial domain by inverse DCT (IDCT) and calculate the saliency map as:

\begin{equation}
m=\operatorname{I D C T}(\hat{I}) \circ \operatorname{I D C T}(\hat{I}),
\end{equation}
where $\circ$ indicates the Hadamard product. As can be seen from Fig. \ref{fig1}, instead of using visual saliency as weighting maps in most existing works, the downsampled saliency maps serve as the reduced-reference information in our proposed framework, which can alleviate the large transmission data of original point clouds.

\begin{figure}[t]
	\centerline{\includegraphics[width=9.0cm]{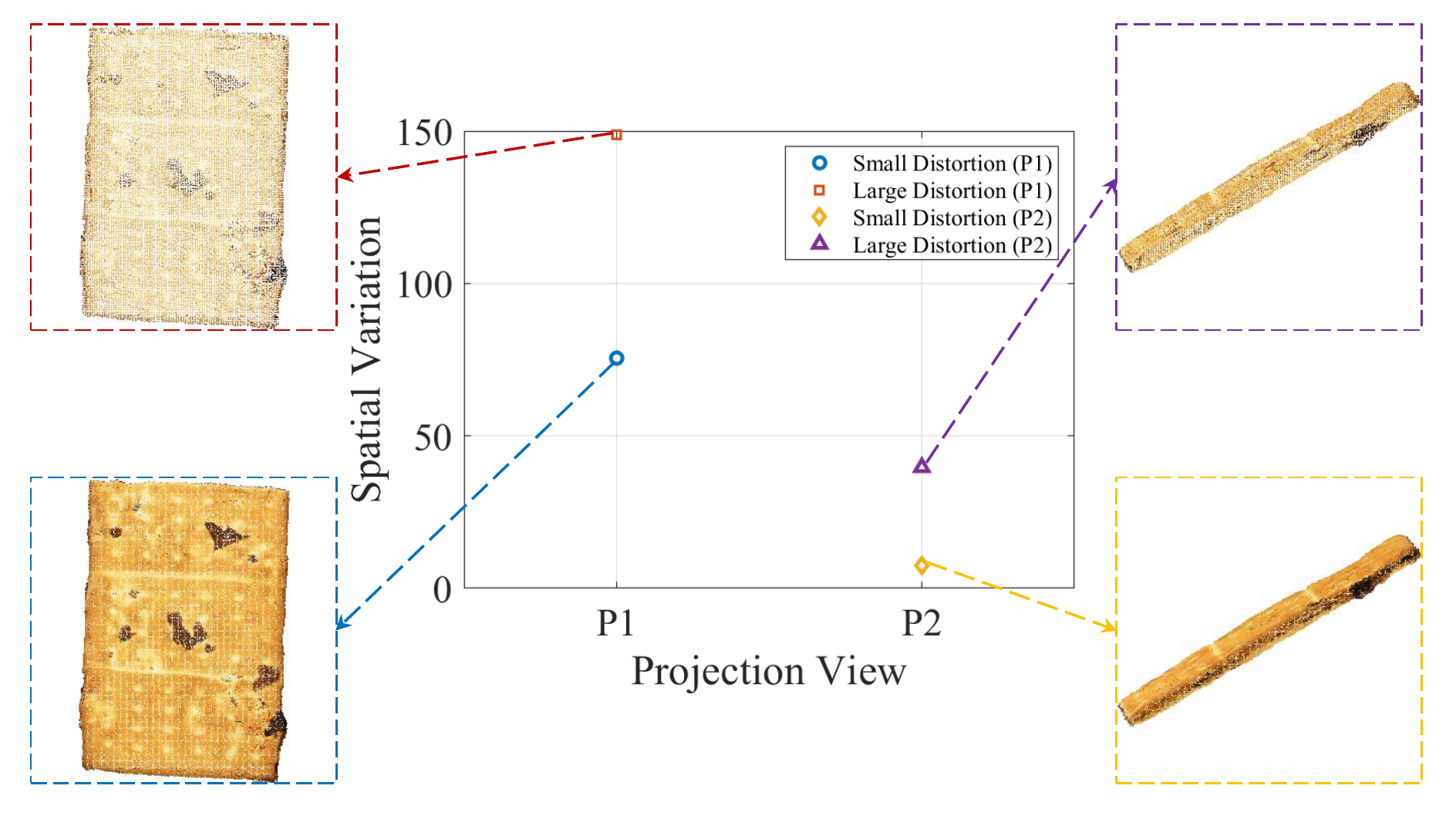}}
	\caption{An example to illustrate the relationship between visual content and quality degradation, where P1 and P2 are two projection views.}
	\centering
	\label{fig2}
\end{figure}

\subsection{Content-Oriented Similarity Measurement}
Intuitively, we can measure image structural similarity between original and test projected saliency maps. Given a distorted saliency map $m_d$ and the corresponding reference saliency map $m_r$, we compute the structural similarity between $m_d$ and $m_r$ as follows:

\begin{equation}
S=\frac{\left(2 \mu_r \mu_d+C_1\right)\left(2 \sigma_{r d}+C_2\right)}{\left(\mu_r^2+\mu_d^2+C_1\right)\left(\sigma_r^2+\sigma_d^2+C_2\right)},
\end{equation}
where $\mu_r$, $\mu_d$, $\sigma_r^2$, $\sigma_d^2$, and $\sigma_{r d}$ represent the corresponding local mean, variance, and covariance of original reference and distorted saliency maps. $C_1$ and $C_2$ are stabilizing constants.

Along with pooling the similarities for all views, we employ a content-oriented weighting strategy, which shows more consistent with the HVS perception. Considering that content is usually quantified by spatial information by using the Sobel filter \cite{yu2013image}, here the spatial variation is taken as the weight and calculated by:

\begin{equation}
w=\left|\operatorname{std}\left[\operatorname{Sobel}\left(I_d\right)\right]-\operatorname{std}\left[\operatorname{Sobel}\left(I_r\right)\right]\right|,
\end{equation}
where $\rm{std[\cdot]}$ denotes the standard deviation that operates over the pixels in the image space. $I_r$ and $I_d$ are the reference and distorted projected images, respectively. Let $n$ be the number of projection views. The content-oriented similarity measurement is obtained as follows:

\begin{equation}
S_w=\frac{1}{n} \sum S^w.
\end{equation}

To show how the proposed content-oriented weighting strategy works, we provide two pairs of projection views with small and large distortion degrees, as illustrated in Fig. \ref{fig2}. We can observe that the visual contents from various views generally have different spatial variations. Additionally, the projection view with large distortion causes more spatial variation.

\subsection{Statistical Correlation Measurement}
Apart from the content-oriented similarity measurement, statistical information from the saliency maps is also important for the perceptual quality of point clouds. Therefore, for each test point cloud, we formulate the statistical correlation measurement by:

\begin{equation}
H_c=\frac{1}{n} \sum \frac{\mathrm{E}\left[h_r h_d\right]-\mathrm{E}\left[h_r\right] \mathrm{E}\left[h_d\right]}{\sqrt{\mathrm{E}\left[h_r^2\right]-\left(\mathrm{E}\left[h_r\right]\right)^2} \sqrt{\mathrm{E}\left[h_d{ }^2\right]-\left(\mathrm{E}\left[h_d\right]\right)^2}},
\end{equation}
where $\mathrm{E}[\cdot]$ represents the expectation operator. $h_r$ and $h_d$ are the statistical histograms of original reference and distorted saliency maps, respectively.

Through the aforementioned procedures, we have two quality measurements, i.e., $S_w$ and $H_c$, for content-oriented similarity and statistical correlation, respectively. Finally, the objective quality scores of distorted point clouds can be calculated as:

\begin{equation}
Q=S_w \cdot H_c.
\end{equation}

\section{Results and Analyses}
In this section, we compare our proposed RR-CAP with state-of-the-art quality assessment methods on public subject-rated quality databases for point clouds. Besides, the ablation study is also conducted to further validate the performance of each proposed technical component.

\subsection{Experimental Protocols}
We carry out extensive experiments on the SJTU-PCQA \cite{yang2020predicting} and WPC \cite{liu2022perceptual} point cloud quality databases. Both databases provide mean opinion score (MOS) as ground-truth subjective quality for each distorted point cloud sample.
\begin{itemize}
    \item SJTU-PCQA database: is composed of 9 reference point cloud samples. Several distortion types including geometry Gaussian noise, color noise, downscaling, octree-based compression, and their mixture are applied to generate 378 distorted point clouds. Each distortion type involves six levels.
    \item WPC database: has 20 original point clouds with four distortion types, resulting in 740 distorted point clouds. To be specific, there exist 60 downsampling distorted point clouds, 180 distorted point clouds with Gaussian noise, 320 G-PCC (T) compressed point clouds, and 180 V-PCC compressed point clouds.
\end{itemize}

Four widely used evaluation criteria are adopted for performance comparisons, including Spearman rank-order correlations coefficient (SROCC), Kendall rank-order correlation coefficient (KROCC), Pearson linear correlation coefficient (PLCC), and root mean square error (RMSE). Among them, SROCC and PLCC/RMSE are employed to measure prediction monotonicity and accuracy, respectively. The KROCC is used for the ordinal association between two measured quantities. An excellent quality metric should obtain the performance values of SROCC, KROCC, and PLCC near 1, and the value of RMSE closes to 0. Note that before computing the PLCC and RMSE for different quality metrics, a nonlinear fitting process \cite{rohaly2000video} is executed to map the predicted quality scores onto the common scale space of subjective scores. 


\begin{table*}[t]
	\begin{center}
		\caption{Performance comparisons of objective quality metrics on SJTU-PCQA and WPC databases. The best performance values for FR, NR, and RR are in bold.}
		\label{table1}
		\scalebox{1.01}{
\begin{tabular}{l|l|l|llll|llll}
\hline
\multirow{2}{*}{Ref} & \multirow{2}{*}{Type} & \multirow{2}{*}{Method} & \multicolumn{4}{c|}{SJTU-PCQA Database} & \multicolumn{4}{c}{WPC Database} \\ \cline{4-11}
& & & SROCC & KROCC & PLCC & RMSE & SROCC & KROCC & PLCC & RMSE \\ \hline
\multirow{5}{*}{FR} &\multirow{3}{*}{Model-based} & GraphSIM \cite{yang2020inferring} & 0.8483 & 0.6448 & 0.8449 & 1.5721 & 0.5831 & 0.4194 & 0.6163 & 17.1939 \\
    & & PointSSIM \cite{alexiou2020towards} & 0.6867 & 0.4964 & 0.7136 & 1.7001 & 0.4542 & 0.3278 & 0.4667 & 20.2733 \\
    & & PCQM \cite{meynet2020pcqm} & \textbf{0.8544} & \textbf{0.6586} & \textbf{0.8653} & \textbf{1.2162} & \textbf{0.7434} & \textbf{0.5601} & \textbf{0.7499} & \textbf{15.1639} \\ \cline{2-11}
    & \multirow{2}{*}{Image-based} & PSNR & 0.2422 & 0.1077 & 0.2317 & 2.3124 & 0.4235 & 0.3080 & 0.4872 & 15.8133 \\
    & & SSIM \cite{wang2004image} & 0.2987 & 0.1919 & 0.3476 & 2.1770 & 0.3878 & 0.3234 & 0.4944 & 15.7749 \\ \hline
\multirow{3}{*}{NR} & Model-based & NR-3DQA \cite{zhang2022no} & \textbf{0.7144} & \textbf{0.5174} & \textbf{0.7382} & \textbf{1.7686} & \textbf{0.6479} & \textbf{0.4417} & \textbf{0.6514} & \textbf{16.5716} \\ \cline{2-11}
    & \multirow{2}{*}{Image-based} & BRISQUE \cite{mittal2012no} & 0.2051 & 0.1121 & 0.2241 & 2.2428 & 0.3781 & 0.2444 & 0.4176 & 22.5414 \\
    & & NIQE \cite{mittal2012making} & 0.2214 & 0.1512 & 0.3764 & 2.2671 & 0.3887 & 0.2551 & 0.3957 & 22.5502 \\ \hline
\multirow{2}{*}{RR} & Model-based & PCMRR \cite{viola2020reduced} & 0.4816 & 0.3362 & 0.6191 & 1.9342 & 0.3097 & 0.2082 & 0.3433 & 21.5302 \\ \cline{2-11}
    & Image-based & \textbf{Proposed RR-CAP} & \textbf{0.7577} & \textbf{0.5508} & \textbf{0.7691} & \textbf{1.5512} & \textbf{0.7162} & \textbf{0.5260} & \textbf{0.7307} & \textbf{15.6485} \\
\hline
\end{tabular}}
	\end{center}
\end{table*}

\subsection{Performance Comparisons}
To verify the performance of our proposed RR-CAP method, we compare it with existing objective quality evaluation approaches including FR, NR, and RR metrics. The results are shown in TABLE \ref{table1}. In general, the compared metrics can be categorized into two types: model-based and image-based methods. The model-based metrics directly operate from 3D models, which consist of GraphSIM \cite{yang2020inferring}, PointSSIM \cite{alexiou2020towards}, PCQM \cite{meynet2020pcqm}, NR-3DQA \cite{zhang2022no}, and PCMRR \cite{viola2020reduced}. The image-based metrics project the 3D models into 2D image planes, and then perform quality assessment on the corresponding projections. The compared image-based metrics involve PNSR, SSIM \cite{wang2004image}, BRISQUE \cite{mittal2012no}, and NIQE \cite{mittal2012making}. Note that PCMRR is the only RR method and belongs to model-based metrics. That is, there has been no investigation on image-based RR metrics specifically designed for point clouds yet. Therefore, we try to fill this gap and propose an image-based RR point cloud metric based on the characteristics of the HVS.

From the quantitative table, we make the following analyses. First, traditional image-based metrics show fewer effects in estimating perceptual quality since they do not take the properties of point clouds into account. Second, our proposed RR-CAP outperforms state-of-the-art NR and RR metrics including both model-based and image-based ones. This demonstrates the superiority of the proposed method although it is an image-based metric. Note that our proposed metric significantly performs better than PCMRR. This is mainly due to the consideration of the HVS perception and viewing process in our framework. Third, we find that the proposed RR metric can greatly reduce the performance gap compared to the FR model-based quality assessment methods.

\subsection{Ablation Tests}
\begin{figure}[t]
	\centerline{\includegraphics[width=9.0cm]{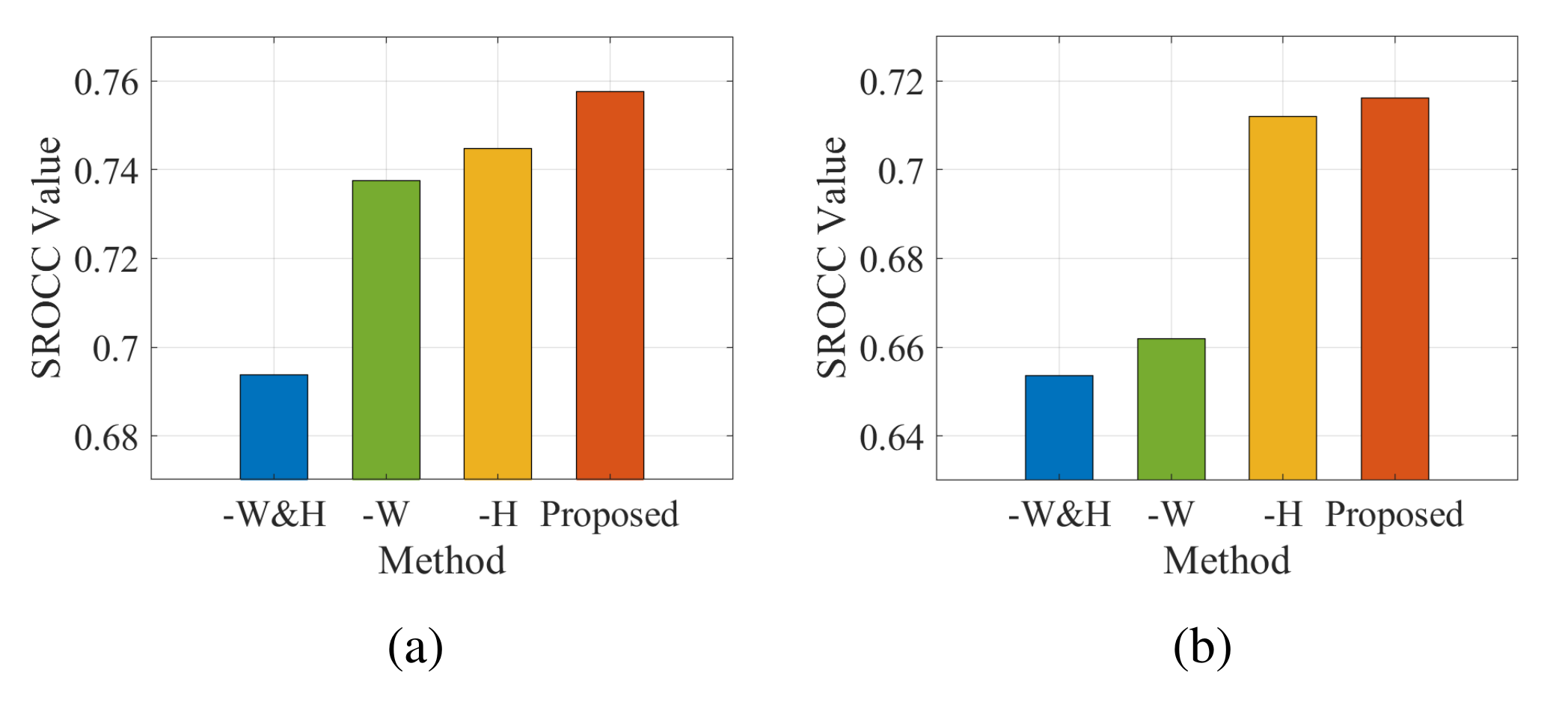}}
	\caption{Performance variation regarding each technical component, where ``-'' indicates removing the corresponding item. W and H are the proposed weighting strategy and statistical histogram feature, respectively. (a) Run on SJTU-PCQA database; (b) Run on WPC database.}
	\centering
	\label{fig3}
\end{figure}

\begin{figure}[t]
	\centerline{\includegraphics[width=9.0cm]{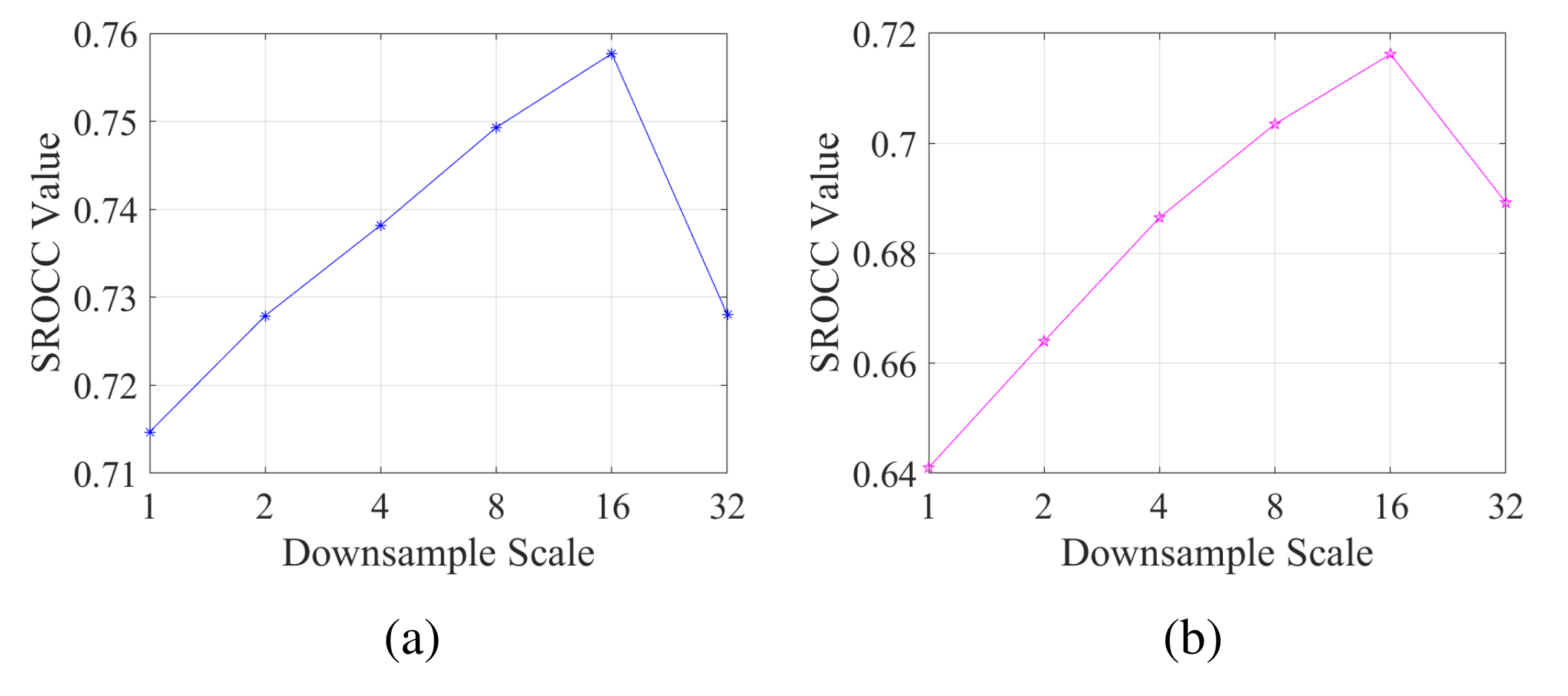}}
	\caption{Performance change with respect to downsampling scales. (a) Run on SJTU-PCQA database; (b) Run on WPC database.}
	\centering
	\label{fig4}
\end{figure}

It is interesting to test the performance variation of each proposed technical component and parameter of our method. In Fig. \ref{fig3} and Fig. \ref{fig4}, we show the tests regarding the content-oriented weighting strategy, statistical histogram feature, and downsampling scales.

From the figures, we can see that our method improves gradually by adding the weighting strategy and statistical information. This validates the effectiveness of the proposed technical components. In addition, larger downsampling scales can save more transmitted resources. To obtain the best performance and also relieve transmission burden, we choose the downsampling scale equaling to $16$ in the proposed framework.

\section{Conclusion}
In this letter, we propose a new general-purpose RR metric for objective quality assessment of point clouds. Inspired by the characteristics of the HVS, our proposed method is based on saliency projection, followed by content-oriented similarity and statistical correlation measurements. Experiments show that the proposed RR-CAP can achieve high consistency with the subjective ratings, compared to state-of-the-art quality metrics. An implementation of the metric is available at: https://github.com/weizhou-geek/RR-CAP.

\bibliographystyle{IEEEtran}
\bibliography{references}

%

%
%
%




\end{document}